\documentclass[aps,pra,showpacs,amsmath,amssymb,amsfonts,superscriptaddress,twocolumn,lengthcheck,longbibliography]{revtex4-2}
\usepackage{amsmath}
\usepackage{graphicx}
\usepackage{subfigure}
\usepackage{verbatim}
\usepackage{dcolumn}
\usepackage{bm}
\usepackage{epsf}
\usepackage{xcolor}
\usepackage{hyperref}
\usepackage{hhline}
\usepackage{txfonts}
\usepackage{float}
\usepackage{enumerate}
\usepackage{bbm}
\usepackage{mathrsfs}
\usepackage{lipsum}
\usepackage{appendix}
\usepackage{braket}
\usepackage{float}
\usepackage{tikz}

\allowdisplaybreaks[1]  

\definecolor{mygreen}{rgb}{0.01, 0.31, 0.59}
\definecolor{myblue}{rgb}{0.01, 0.31, 0.59}
\hypersetup{
	colorlinks=true,   
	linkcolor=blue,  
	citecolor=blue, 
	urlcolor =blue    
}

\begin{document}
	
	\title{Macroscopic entanglement between two magnon modes via two-tone driving \\ of a superconducting qubit} 
	
	\author{Rong-Can Yang} 
	\email{\textcolor{black}{These authors contributed equally to this work.}}
	\affiliation{Zhejiang Key Laboratory of Micro-Nano Quantum Chips and Quantum Control, 	and School of Physics, Zhejiang University, Hangzhou 310027, China}
	\affiliation{Fujian Provincial Key Laboratory of Quantum Manipulation and New Energy Materials, College of Physics and Energy, Fujian Normal University, Fuzhou, 350117, China}
	\affiliation{Fujian Provincial Engineering Technology Research Center of Solar Energy Conversion and Energy Storage, Fuzhou, 350117, China}
	
	\author{Gang Liu}
	\email{\textcolor{black}{These authors contributed equally to this work.}}
	\affiliation{Zhejiang Key Laboratory of Micro-Nano Quantum Chips and Quantum Control,
		and School of Physics, Zhejiang University, Hangzhou 310027, China}
	
	\author{Gen Li}
	\affiliation{Zhejiang Key Laboratory of Micro-Nano Quantum Chips and Quantum Control,
		and School of Physics, Zhejiang University, Hangzhou 310027, China}
	
	\author{Jie Li}
	\email[Contact author: ]{jieli007@zju.edu.cn}
	\affiliation{Zhejiang Key Laboratory of Micro-Nano Quantum Chips and Quantum Control,
		and School of Physics, Zhejiang University, Hangzhou 310027, China}	
	
	\date{\today}

\begin{abstract}
	The cavity-mediated coupling between magnons in an yttrium-iron-garnet (YIG) sphere and a superconducting qubit has recently been demonstrated as a new platform for preparing macroscopic quantum states. Here, based on this system, we propose to entangle two magnon modes in two YIG spheres by driving the qubit with a two-tone field and by appropriately choosing the frequencies and strengths of the two driving fields. We show that strong entanglement can be achieved with fully feasible parameters.  We further provide a detection scheme for experimentally verifying the entanglement. Our results indicate that macroscopic entanglement between two magnon modes in two millimeter-sized YIG spheres, involving more than $10^{18}$ spins, can be realized using currently available parameters, which finds promising applications in fundamental studies, such as macroscopic quantum mechanics and the test of unconventional decoherence theories.
	\end{abstract}

	\maketitle

	\section{Introduction}
	
    In recent years, hybrid quantum systems based magnons in yttrium iron garnet (YIG) have made significant progress in both theory and experiment~\cite{Lachance2019,YUAN20221,Zare2022,Zuo_2024}.  In particular, the realization of strong coupling between microwave cavity photons and magnons in a YIG sphere~\cite{PhysRevLett.111.127003,Tabuchi2014,PhysRevLett.113.156401,PhysRevApplied.2.054002,PhysRevLett.114.227201} has greatly stimulated the research in the fields of cavity, nonlinear, and quantum magnonics. A growing number of studies have indicated the great potential of this novel magnonic hybrid system in quantum information processing, quantum technology, quantum sensing, and fundamental studies, e.g., macroscopic quantum mechanics. An important reason is that the YIG sphere has, so far, the smallest dissipation rate for the magnon mode, compared to other ferromagnetic and antiferromagnetic materials~\cite{Serha2025}, which is vital for activating any real quantum applications and preparing magnonic quantum states. Another distinct advantage of the system is manifested as its excellent ability to coherently couple with various quantum systems, e.g., microwave photons~\cite{PhysRevLett.111.127003,Tabuchi2014,PhysRevLett.113.156401,PhysRevApplied.2.054002,PhysRevLett.114.227201}, optical photons~\cite{PhysRevB.93.174427,PhysRevLett.116.223601,PhysRevLett.117.123605,PhysRevLett.117.133602}, phonons~\cite{PhysRevLett.121.203601,ZhangXF2016, PhysRevX.11.031053,PhysRevLett.129.123601,PhysRevLett.129.243601,Shen2025}, and superconducting qubits~\cite{Tabuchi2015,Lachance2017,Lachance2020,Xuda2023,Xuda2026}. In particular, the magnon--superconducting--qubit system via the mediation of a microwave cavity has now became a new platform for preparing macroscopic quantum states~\cite{Lachance2020,Xuda2023,Xuda2026}. Significant experimental breakthroughs include the realization of magnon--qubit strong coupling~\cite{Tabuchi2015}, single-shot detection of a single magnon~\cite{Lachance2020}, superposition states of a single magnon~\cite{Xuda2023}, and magnon squeezing in the quantum regime~\cite{Xuda2026}. In the meantime, many theoretical work have proved the outstanding ability of this system in preparing magnonic quantum states, including magnon blockade~\cite{Liuzx2019,PhysRevA.101.042331,PhysRevA.101.063838,PhysRevA.103.052411,PhysRevB.104.224434,WangYM2022,PhysRevA.109.023710}, squeezed~\cite{PhysRevA.108.063703,PhysRevA.111.053707,Liugang2026}, cat~\cite{Kounalakis2022,PhysRevA.107.023709,PhysRevA.110.013711,PhysRevA.110.053710,Liugang2025}, and entangled~\cite{PhysRevB.103.224416,Luo:21,Qisf2022,Renyl2022} states. 
	
    Among various quantum states, magnon entanglement involving macroscopic YIG spheres is of particular interest, since it is a genuine macroscopic quantum state related to many fundamental issues of quantum mechanics, such as the Schr\"odinger's cat paradox, the quantum-to-classical transition, and decoherence theories. In the past decade, a number of proposals have been offered to prepare magnon entanglement of two YIG spheres by exploiting, e.g., the magnetostrictive interaction~\cite{Li_2019}, magnon Kerr effect~\cite{PhysRevResearch.1.023021}, squeezed microwave fields~\cite{Yu_2020,Nair2020,Yangzb2021,PhysRevA.110.063715,7hhn-zypy}, optomagnonic Brillouin light scattering~\cite{WuWJ2021,HuNQ2025}, two-photon~\cite{PhysRevB.106.224404} and floquet~\cite{Xie2023,Hu:24} driving, parity measurement~\cite{Yan2024}, etc. Besides, establishing entanglement of multiple spheres is also of considerable interest~\cite{PhysRevA.111.033712}.  It is worth mentioning that proposals have also been provided to entangle magnons in antiferromagnetic systems~\cite{PhysRevB.101.014419,ZhengSS2020,PhysRevB.104.224302}, in which magnons suffer much greater dissipation. 
    
   Here, we propose to entangle two magnon modes of two YIG spheres by applying a two-tone driving field to a superconducting qubit that is effectively coupled to the two magnon modes via a microwave cavity. Specifically, the qubit and the two YIG spheres are placed inside and coupled to a common three-dimensional microwave cavity.  The cavity is far-detuned from both the magnon modes and the qubit, such that the cavity is only virtually excited and establishes an effective magnon--qubit and magnon--magnon coupling.     
   By appropriately modulating the two driving fields, in which one is resonant with the qubit with large driving strength to induce an effective Rabi-type magnon--qubit coupling and the frequency and strength of the other drive are precisely chosen, an effective magnonic two-mode squeezing interaction can be derived, which yields entanglement between the two magnon modes. 
   Numerical simulations based on the master equation indicate that considerable entanglement is achievable using currently available parameters. An entanglement detection scheme is also discussed by performing a joint Wigner-function tomography of the two magnon modes exploiting their dispersive coupling to the qubit.
   {Compared to the existing protocols mentioned above, the main advantage of this protocol lies in its high feasibility, i.e., the superconducting qubit can provide sufficient strong nonlinearity to create entanglement, while this remains challenging for the protocols based on nonlinear magnomechanical~\cite{Li_2019} and optomagnonic interaction~\cite{WuWJ2021,HuNQ2025}, and magnon Kerr effect~\cite{PhysRevResearch.1.023021}. Moreover, our protocol does not require external quantum driving fields~\cite{Yu_2020,Nair2020,Yangzb2021,PhysRevA.110.063715,7hhn-zypy,PhysRevB.106.224404}, as well as modulated driving fields, e.g., Floquet engineering~\cite{Xie2023,Hu:24}. 
Our protocol also does not need a second cavity mode and the second-excited state of the transmon with respect to the protocol of preparing magnon Bell states~\cite{Yan2024}. }
     
   The remainder of the paper is organized as follows. In Sec.~\ref{sec:2}, we introduce the cavity--magnon--qubit system and show how to derive an effective two-mode squeezing interaction between two magnon modes by adiabatically eliminating the cavity and appropriately choosing the frequencies and strengths of the two driving fields applied to the qubit. In Sec.~\ref{sec:3}, we derive an effective master equation for numerical simulation, and present the results of the magnonic entanglement. We further provide an entanglement detection scheme for experimental verification. Finally, we summarize our findings in Sec.~\ref{sec:4}.

	\section{The system and effective Hamiltonian \label{sec:2}}

	\begin{figure}
		\centering
		\includegraphics[width=\linewidth]{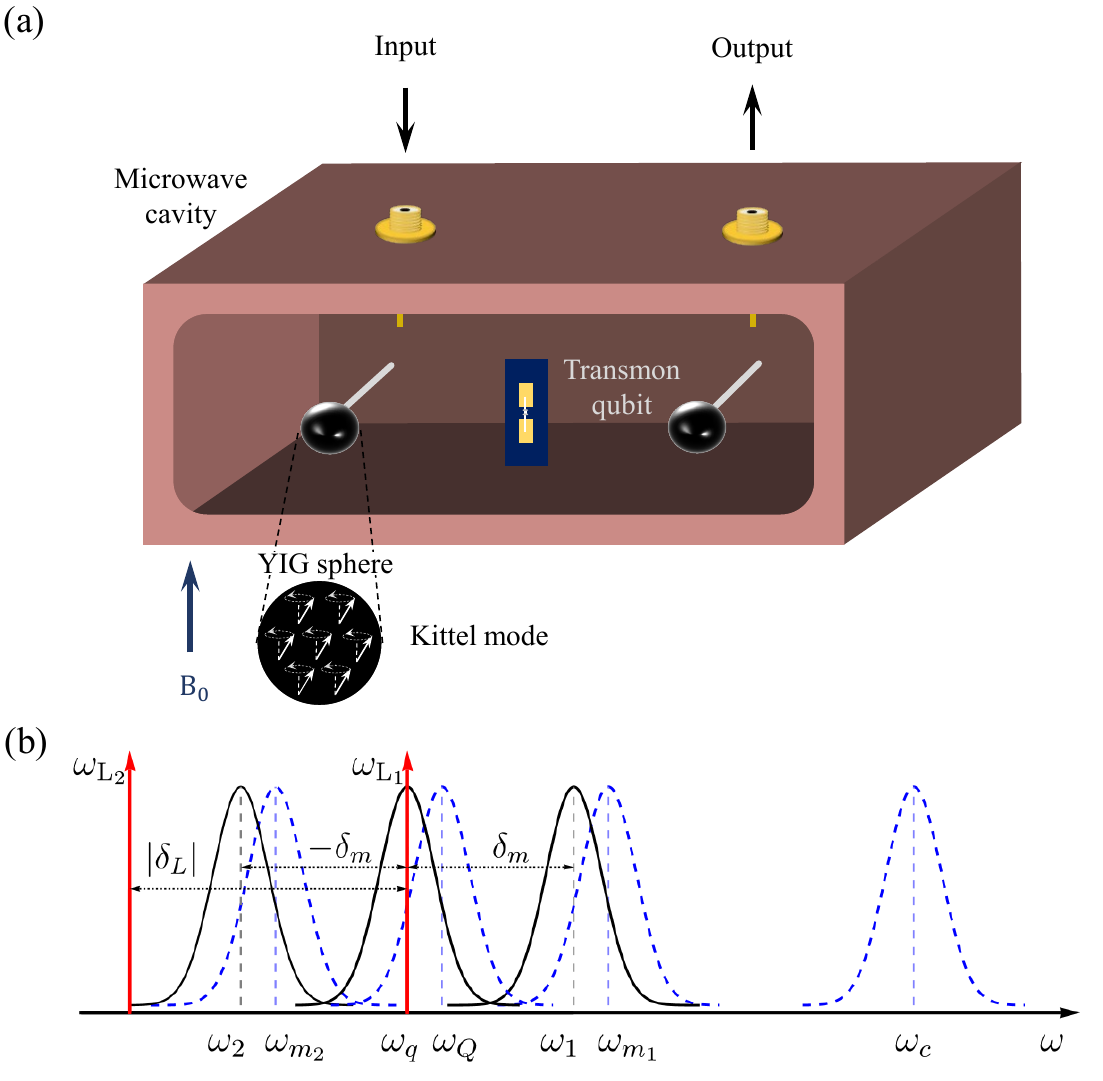}
		\caption{(a) The cavity--magnon--qubit system used for generating macroscopic entanglement between two magnon modes. Two magnon modes in two YIG spheres and a transmon superconducting qubit couple to a common microwave cavity mode via magnetic- and electric-dipole interactions, respectively. The two YIG spheres are placed inside a uniform bias magnetic field $B_0$, and each sphere is further in a local magnetic field provided by a small coil (not shown). The qubit is driven by two microwave fields. (b) Original and effective frequencies of the system with two qubit drive frequencies. The cavity mode at frequency $\omega_c$ is far-detuned from both the magnon modes at $\omega_{m_1}$ and $\omega_{m_2}$ and the qubit at $\omega_Q$. Adiabatic elimination of the cavity mode yields the magnon effective frequencies $\omega_1$ and $\omega_2$, and the qubit effective frequency $\omega_q$. The qubit is driven by two microwave fields at frequencies $\omega_{\text{L}_1}$ and $\omega_{\text{L}_2}$. See text for details about the frequencies and strengths of the two driving fields.}
		\label{fig1}
	\end{figure}

		We consider a hybrid cavity--magnon--qubit system~\cite{Tabuchi2015,Lachance2017,Lachance2020,Xuda2023,Xuda2026}, where two millimeter-sized YIG spheres and a transmon-type superconducting qubit are placed inside a microwave cavity, as illustrated in Fig.~\ref{fig1}. 	Each YIG sphere supports a magnon mode, e.g., the Kittel mode, and is positioned inside a uniform static magnetic field $B_0$. In addition, a small direct-current coil provides independent fine-tuning of a local magnetic field for each sphere, such that the two magnon frequencies can be independently adjusted within a range required by our protocol. The qubit is simultaneously driven by two coherent microwave fields. 	
	Both the magnon modes and the qubit are strongly coupled to the microwave cavity mode via the magnetic- and electric-dipole interactions, respectively, with coupling rates $g_{\mathrm{c}k}$ $(k=1,2)$ and $g_{\mathrm{cq}}$. The Hamiltonian of the whole system reads (setting $\hbar = 1$)
	\begin{align}\label{Ham_ff}
		H =& \sum_{k=1}^{2}\omega_{m_k}m_k^\dagger m_k + \omega_c c^\dagger c
		+ \frac{\omega_Q}{2}\sigma_z + g_{\mathrm{cq}} \left(c \sigma_+ + c^\dag\sigma_- \right)\nonumber\\
		&+ \left(\sum_{k=1}^{2}g_{\mathrm{c}k} c m_k^\dagger
		+ \sum_{j=1}^{2}\Omega_j \sigma_+ e^{-i\omega_{\text{L}_j}t} + \mathrm{H.c.}\right),
	\end{align}
	where $c$ ($c^\dagger$) and $m_k$ ($m_k^\dagger$) are the annihilation (creation) operators of the cavity mode and the $k$th magnon mode with frequencies $\omega_c$ and $\omega_{m_k}$, respectively. $\omega_Q$ is the qubit transition frequency, with the lowering (raising) operator $\sigma_- = |g\rangle\langle e|$ ($\sigma_+ = |e\rangle\langle g|$) and the Pauli operator 
	$\sigma_z = |e\rangle\langle e| - |g\rangle\langle g|$, and $\Omega_j$ denotes the Rabi frequency associated with the $k$th microwave driving field at frequency $\omega_{\text{L}_j}$ applied to the qubit.

	When the cavity mode is far-detuned from the qubit and the magnon modes, i.e., $\Delta_q \equiv \omega_c - \omega_Q \gg g_{\mathrm{cq}}$ and $\Delta_k \equiv \omega_c - \omega_{m_k} \gg g_{\mathrm{c}k}$, the cavity mode is only virtually excited and can be adiabatically eliminated, which yields an effective Jaynes--Cummings-type interaction between the qubit and each magnon mode and an effective beam-splitter coupling between the two magnon modes. The resulting effective Hamiltonian of the magnon--qubit system is given by
	\begin{align}\label{Ham_full}
		H_1 =& \sum_{k=1}^{2} \omega_k m_k^\dagger m_k + \frac{\omega_q}{2}\sigma_z -  g_{12} \left(m_1^\dagger m_2 + m_1 m_2^\dagger \right)\nonumber\\
		& \left( - \sum_{k=1}^{2}g_{\mathrm{q}k} m_k^\dagger \sigma_-  +  \sum_{j=1}^{2}\Omega_j  \sigma_+ e^{-i \omega_{\mathrm L_j}t} + \text{H.c.} \right),
	\end{align}
	where $\omega_q = \omega_Q - g_{\mathrm{cq}}^2/\Delta_q$ and $\omega_k = \omega_{m_k} - g_{\mathrm{c}k}^2/\Delta_k$ are the modified effective frequencies of the qubit and the $k$th magnon mode, respectively (cf. Fig.~\ref{fig1}(b)), $g_{\mathrm{q}k} = g_{\mathrm{cq}}g_{\mathrm{c}k} \left(\Delta_q^{-1}+\Delta_k^{-1} \right)/2$ is the effective coupling strength between the qubit and the $k$th magnon mode, and $g_{12} = g_{\mathrm{c}1}g_{\mathrm{c}2} \left(\Delta_1^{-1}+\Delta_2^{-1} \right)/2$ is the effective coupling strength between the two magnon modes.

	In the frame rotating at the first drive frequency $\omega_{\text{L}_1}$, 
	the Hamiltonian~\eqref{Ham_full} becomes
	\begin{align}\label{Ham_f1}
		H_2 = & \sum_{k=1}^{2} \delta_k m_k^\dagger m_k +\frac{\delta_q}{2}\sigma_z - g_{12} \left(m_1^\dagger m_2 + m_1 m_2^\dagger \right)   \nonumber\\  
		& + \left[ \left( -\sum_{k=1}^{2} g_{\mathrm{q}k} m_k  + \Omega_1 + \Omega_2 e^{-i \delta_Lt}   \right)\sigma_+ +  \text{H.c.}\right],
	\end{align}
	where $\delta_{q(k)} = \omega_{q(k)} - \omega_{\text{L}_1}$, and $\delta_L = \omega_{\text{L}_2} - \omega_{\text{L}_1}$ is the detuning between the two driving fields. Expressing the qubit operators in the dressed-state basis $\{|+\rangle, |-\rangle\}$ with $|\pm\rangle = \left( |e\rangle \pm |g\rangle \right)/\sqrt{2}$, and introducing the corresponding Pauli operators $\tau_x = \sigma_{+-}+\sigma_{-+} = \sigma_z$, $\tau_y = -i(\sigma_{+-}-\sigma_{-+}) = -\sigma_y$, and $\tau_z = \sigma_{++}-\sigma_{--}=\sigma_x$ with $\sigma_{jk} = |j\rangle\langle k|$ 	$(j,k=\pm)$, the above Hamiltonian is rewritten as
	\begin{align}\label{Hf2}
		H_2 =& \sum_{k=1}^{2}\delta_{k} m_k^\dagger m_k  +  \frac{\delta_q }{2}\tau_x+ \Omega_1 \tau_z - g_{12} \left(m_1^\dagger m_2 + m_1 m_2^\dagger \right) \nonumber\\
		&- \left[ \frac{1}{2}\left(g_{\mathrm q1} m_1 + g_{\mathrm q2} m_2 - \Omega_2 e^{-i\delta_Lt} \right) \left(\tau_z - i\tau_y \right) + \mathrm{H.c.} \right].
	\end{align}
	
	{For simplicity, we set $\delta_q \approx 0$, such that the first drive is nearly resonant with the qubit and creates two dressed states $\ket{\pm}$ with an energy level splitting of $2\Omega_1$.~The qubit then interacts with the two magnon modes via dipole coupling in the dressed-state basis.} Moving to the interaction picture with respect to $\Omega_1\tau_z$, the dressed-state operators transform as $\tau_x \to \tau_x\cos (2\Omega_1 t) - \tau_y\sin (2\Omega_1 t)$, $\tau_y \to \tau_y\cos (2\Omega_1 t) + \tau_x\sin (2\Omega_1 t)$, and $\tau_z \to \tau_z$, which lead the Hamiltonian~\eqref{Ham_f1} to the following form
	\begin{align}\label{Hf3}
		H_3 =& \sum_{k=1}^{2}\delta_{k} m_k^\dagger m_k  
		- \left\{g_{12} m_1^\dagger m_2 \right. + \frac{1}{2}\left(g_{\mathrm q1} m_1 + g_{\mathrm q2} m_2 - \Omega_2 e^{-i\delta_L t} \right) \nonumber\\
		&\times \left. \left[\tau_z - i \left(\tau_y \cos 2\Omega_1t + \tau_x \sin 2\Omega_1 t \right) \right] + \mathrm{H.c.}\right\}.
	\end{align}
{The presence of the second drive with detuning $\delta_L$ and Rabi frequency $\Omega_2$ induces transitions between dressed states with different photon number $n$~\cite{shirley1965solution, cohen1969absorption}.~When $\delta_L = -2\Omega_1$ is chosen, the drive is resonant with the transition between $\ket{+}\ket{n}$ and $\ket{-}\ket{n+1}$.} Under the conditions $|\delta_L| \gg \left\{\Omega_2,\,g_{\mathrm{q}1}/2,\,g_{\mathrm{q}2}/2,\,\delta_1,\,\delta_2\right \}$, the fast-oscillating terms at frequencies $2\Omega_1$ and $4\Omega_1$ can be safely neglected under the rotating-wave approximation (RWA). Consequently, we obtain
	\begin{align}\label{Hqm_pp}
		H_4 =& \sum_{k=1}^{2}\delta_{k} m_k^\dagger m_k  + \frac{\Omega_2}{2}\sigma_z - g_{12}\left(m_1^\dagger m_2 + m_1 m_2^\dagger \right) \nonumber\\
		& - \frac{1}{2}\left(g_{\mathrm q1} m_1 + g_{\mathrm q2} m_2 + \text{H.c.} \right) \sigma_x.
	\end{align}
	Note that we have replaced $\tau_x$ ($\tau_z$) with $\sigma_z$ ($\sigma_x$) in getting Eq.~\eqref{Hqm_pp}.
	By assuming $\delta_1 = -\delta_2 \equiv \delta_m > 0$, {which means that the sum of the two magnon modes frequencies is close to twice the resonance frequency of the qubit}, and working in the interaction picture with respect to $\frac{\Omega_2}{2}\sigma_z + \delta_m (m_1^\dagger m_1 - m_2^\dagger m_2)$, we achieve
	\begin{align}\label{HI}
		H_5 =&\, -g_{12}\,m_1^\dagger m_2\,e^{i2\delta_m t}
		- \frac{1}{2} \left(g_{\mathrm{q}1}m_1\,e^{-i\delta_m t} + g_{\mathrm{q}2}m_2\,e^{i\delta_m t} \right) \nonumber\\
		&\times \left(\sigma_+\,e^{i\Omega_2 t} + \sigma_-\,e^{-i\Omega_2 t} \right)
		+ \mathrm{H.c.}
	\end{align}
	
	Under the conditions $\Omega_2 > \delta_m$ and $\{\Omega_2 \pm \delta_m,\, 2\delta_m\} \gg \{2g_{12},\,g_{\mathrm{q}1},\,g_{\mathrm{q}2}\}/2$, {direct energy exchanges between the qubit and the two magnon modes are suppressed. {However, via a second-order virtual process mediated by the qubit, the second drive enables the simultaneous creation or annihilation of one magnon in each of the two magnon modes.}} After that, the effective Hamiltonian of the system is (see Appendix~\ref{appendix} for details)
	\begin{align}\label{Heff0}
		H_{\mathrm{eff}} 
			=&\ \frac{g_{12}^2}{2\delta_m}(n_1-n_2)\nonumber\\
			+&\ \frac{1}{4(\Omega_2+\delta_m)}
				\Bigl[\bigl(g_{\mathrm{q}1}^2 n_1+g_{\mathrm{q}2}^2 n_2\bigr)
				+g_{\mathrm{q}1}g_{\mathrm{q}2}
				\bigl(m_1^\dagger m_2^\dagger+m_1 m_2\bigr)\Bigr]\sigma_z
				\nonumber\\
			+&\ \frac{1}{4(\Omega_2-\delta_m)}
				\Bigl[\bigl(g_{\mathrm{q}1}^2 n_1+g_{\mathrm{q}2}^2 n_2\bigr)
				+g_{\mathrm{q}1}g_{\mathrm{q}2}
				\bigl(m_1^\dagger m_2^\dagger+m_1 m_2\bigr)\Bigr]\sigma_z.\nonumber\\
			+&\{\xi_1[g_{12}m_1^\dagger m_2,(g_{\text{q}1}m_1+g_{\text{q}2}m_2^\dagger)\sigma_-]e^{-i(\Omega_2 - \delta_m)t}\nonumber\\
			&\ +\xi_2[g_{12}m_1^\dagger m_2,(g_{\text{q}1}m_1^\dagger+g_{\text{q}2}m_2)\sigma_-]e^{-i(\Omega_2 - 3\delta_m)t}\nonumber\\
			&\ +\xi_3[(g_{\mathrm{q}1} m_1 + g_{\mathrm{q}2} m_2^\dagger)\sigma^+,(g_{\text{q}1}m_1+g_{\text{q}2}m_2^\dagger)\sigma_-]e^{-i2\delta_mt}\nonumber\\
			&\ +\mathrm{H.c.}\},
	\end{align}			
		where $n_i\equiv m_i^\dagger m_i(i = 1,2)$, $\xi_1 = (\Omega_2 + 3\delta_m)/[8\delta_m(\Omega_2 + \delta_m)]$, $\xi_2 = (\Omega_2 + \delta_m)/[8\delta_m(\Omega_2 - \delta_m)]$, and $\xi_3 = \Omega_2/[4(\Omega_2^2 - \delta_m^2)]$.~If the further condition $\{\Omega_2 - \delta_m,|\Omega_2 - 3\delta_m|,2\delta_m \} \gg \{\xi_1g_{12}g_{\text{q}k},\xi_2g_{12}g_{\text{q}k},\xi_3 g_{\text{q}k}g_{\text{q}l}\}$ ($k,l\in\{1,2\}$) is satisfied, the terms oscillating at frequencies $\Omega_2-\delta_m$, $|\Omega_2-3\delta_m|$, and $2\delta_m$ in Eq.~\eqref{Heff0} can be neglected under a RWA. 		
		By evaluating the first three time-independent terms, we arrive at the effective Hamiltonian
		\begin{align}\label{Heff}
			H_{\mathrm{eff}} \approx&\,\frac{g_{12}^2}{2\delta_m} (m_1^\dagger m_1 - m_2^\dagger m_2 ) + [r_g  (g_{\mathrm q1}^2m_1^\dagger m_1 + g_{\mathrm q2}^2m_2^\dagger m_2 ) \nonumber\\
			&+ g_\text{eff} (m_1^\dagger m_2^\dagger + m_1 m_2 )]\sigma_z,
		\end{align}
		where $g_{\mathrm{eff}} = r_gg_{\mathrm{q}1}g_{\mathrm{q}2}$ is the effective coupling strength of the magnonic two-mode squeezing interaction, with $r_g = \big[(\Omega_2-\delta_m)^{-1} + (\Omega_2+\delta_m)^{-1} \big]/4$.  Clearly, the last term $g_{\mathrm{eff}}(m_1^\dagger m_2^\dagger + m_1 m_2)\sigma_z$ gives rise to entanglement between two magnon modes, no matter the qubit is initialized in the ground or excited state.

	\section{Magnonic entanglement and its detection} \label{sec:3}

	In Sec.~\ref{sec:2}, we have analytically derived the effective Hamiltonian for magnon entanglement without considering any dissipation and dephasing of the system.  To account for these effects in a practical situation, we adopt the Lindblad master equation to solve the density matrix $\rho$ of the magnon--qubit system, from which the reduced density matrix of two magnon modes can be achieved by tracing out the qubit degree of freedom, i.e., $\rho_{12} = \mathrm{tr}_q \, \rho$.  Since the effective Hamiltonian~\eqref{Heff} is derived from the Hamiltonian~\eqref{Ham_f1} through a sequence of unitary transformations {and approximations}, the master equation must be transformed accordingly. The master equation associated with the Hamiltonian~\eqref{Ham_f1} is given by~\cite{Breuer2007}
	\begin{align}\label{ME}  
		\frac{d\rho}{dt} =& - i [H_2, \rho] + \sum_{k=1}^{2}\frac{\kappa_k(\bar{n}_{k} + 1)}{2} \mathcal{L} [m_k]\rho  + \frac{\kappa_k \bar{n}_{k}}{2} \mathcal{L} [m_k^\dag]\rho \nonumber\\
		&+ \frac{\gamma_d (\bar{n}_{q} + 1)}{2} \mathcal{L} [\sigma_-]\rho + \frac{\gamma_d \bar{n}_{q}}{2} \mathcal{L} [\sigma_+]\rho + \frac{\gamma_{\phi}}{4} \mathcal{L} [\sigma_z]\rho, 
	\end{align}
	where $\mathcal{L}[o]\rho = 2o\rho o^\dagger - (o^\dagger o\rho + \rho o^\dagger o)$
	is the Lindblad superoperator, $\kappa_k$ is the damping rate of the $k$th magnon
	mode, $\gamma_d$ ($\gamma_\phi$) is the qubit dissipation (dephasing) rate, and
	$\bar{n}_{k(q)} = \{\exp[\hbar\omega_{m_k(q)}/k_BT]-1\}^{-1}$ is the mean thermal
	excitation number of the $k$th magnon mode (qubit), with the bath temperature $T$ and
	the Boltzmann constant $k_B$. 
	Applying the same unitary transformations {and approximations} in Sec.~\ref{sec:2} leads to the following effective master equation associated with the effective Hamiltonian~\eqref{Heff}:
	\begin{eqnarray}\label{EME} 
		\frac{d\tilde\rho}{dt} &=& - i \left[H_\text{eff}, \tilde\rho\right] + \sum_{k=1}^{2} \frac{\kappa_k (\bar{n}_{k} + 1 )}{2} \mathcal{L}[m_k]\tilde\rho + \frac{\kappa_k \bar{n}_{k}}{2} \mathcal{L} [m_k^\dag]\tilde\rho \nonumber \\
		&+& \xi_1 \big(\mathcal{L} [\sigma_+]\tilde\rho + \mathcal{L} [\sigma_-]\tilde\rho \big) + \xi_2 \mathcal{L} [\sigma_z]\tilde\rho,
	\end{eqnarray}
	where $\tilde\rho$ is the accordingly transformed density matrix,  and $\xi_1 = \big[3\gamma_d(2\bar{n}_{q} + 1 ) + 2\gamma_{\phi} \big]/16$ and $\xi_2 = \big[\gamma_d(2\bar{n}_{q} + 1 ) + 2\gamma_{\phi} \big]/16$.

	We assume the system is initialized in the state $|g00\rangle$, i.e., the qubit in the ground state and the two magnon modes in the vacuum state, which can be achieved by simply placing the system at a low temperature, e.g.,10 mK. For the qubit and magnon at GHz frequencies, this gives vanishing thermal excitations $\bar{n}_{m_k(q)}  \approx 0$. Our protocol generates dynamical entanglement between two magnon modes, and the entanglement is quantified by the logarithmic negativity (LN)~\cite{PhysRevA.58.883,PhysRevA.65.032314}. 
	For a two-mode magnonic state $\rho_{12}$,	the LN is defined as $E_N = \ln\|\rho_{12}^{T_k}\|$, where $\rho_{12}^{T_k}$ denotes the partial transpose of $\rho_{12}$ with respect to the $k$th mode and $\|\cdot\|$ signifies the trace norm.  We adopt the following experimentally feasible parameters~\cite{Xuda2023,Xuda2026}: $\omega_c/2\pi = 6.388~\mathrm{GHz}$,	$\omega_Q/2\pi = 5.843~\mathrm{GHz}$, $\omega_{m_1}/2\pi = 5.871~\mathrm{GHz}$,	$\omega_{m_2}/2\pi = 5.801~\mathrm{GHz}$, $g_\mathrm{cq}/2\pi = 80~\mathrm{MHz}$, $g_{\mathrm{c}1}/2\pi = g_{\mathrm{c}2}/2\pi = 50~\mathrm{MHz}$, corresponding to the detunings $\Delta_q/2\pi = 545~\mathrm{MHz}$, $\Delta_1/2\pi = 517~\mathrm{MHz}$, and $\Delta_2/2\pi = 587~\mathrm{MHz}$. 
	The adiabatic elimination of the cavity leads to the following effective qubit and magnon frequencies: $\omega_q/2\pi = 5.8313~\mathrm{GHz}$, $\omega_{1}/2\pi = 5.8662~\mathrm{GHz}$, and $\omega_{2}/2\pi = 5.7967~\mathrm{GHz}$, and the effective coupling strengths:	$g_{\mathrm{q}1}/2\pi = 7.54~\mathrm{MHz}$, $g_{\mathrm{q}2}/2\pi = 7.08~\mathrm{MHz}$, and $g_{12}/2\pi = 4.55~\mathrm{MHz}$. For two qubit driving fields, we fix $\omega_{\text{L}_1}/2\pi = 5.83145~\mathrm{GHz}$ and $\Omega_2/2\pi = 95~\mathrm{MHz}$, which yield $\delta_m/2\pi  = 34.75~\mathrm{MHz}$ and $\delta_q/2\pi = 0.15~\mathrm{MHz}$, while we treat $\Omega_1$ and thus $\omega_{\text{L}_2} = \omega_{\text{L}_1} -2\Omega_1$ as variables for verifying the conditions assumed in Sec.~\ref{sec:2} for the RWA.	
	
	\begin{figure}[t]
		\centering
		\includegraphics[width=0.95\linewidth,keepaspectratio]{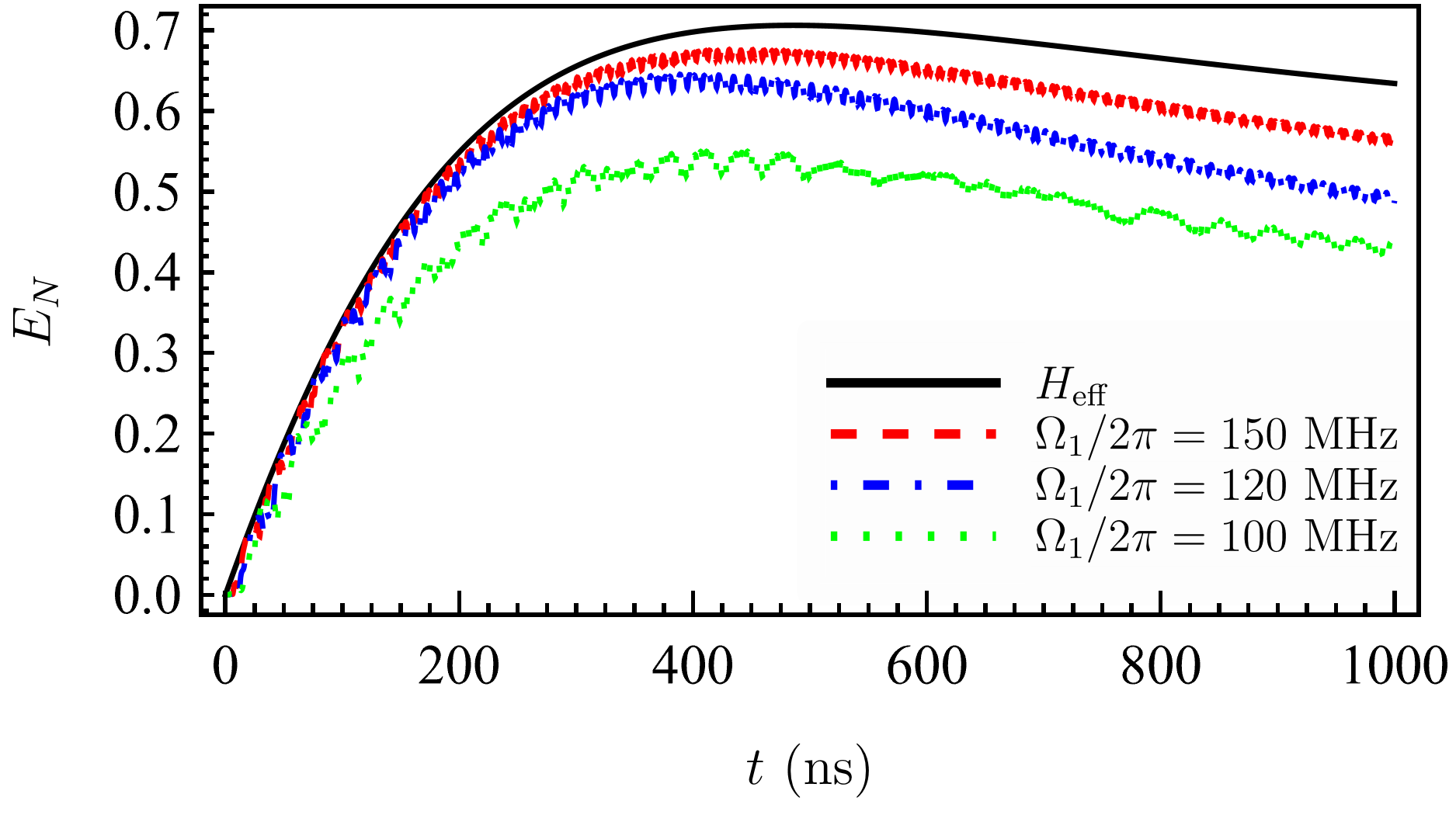}
		\caption{Entanglement between two magnon modes as a function of time.~The black solid curve is obtained from the derived effective Hamiltonian~\eqref{Heff}, while the other three curves are achieved from the Hamiltonian~\eqref{Ham_f1} for three values of $\Omega_1/2\pi = 100, 120$, and 150 MHz.~Other parameters are given in the main text.}
		\label{fig2}
	\end{figure}
	
	In Fig.~\ref{fig2}, we compare the time evolution of the magnon entanglement obtained from the effective magnon--qubit Hamiltonian~\eqref{Heff} with that from the Hamiltonian $H_2$ in Eq.~\eqref{Ham_f1} after the elimination of the cavity. We consider currently available dissipation and dephasing rates: $\kappa_{1,2}/2\pi = 0.5~\mathrm{MHz}$~\cite{Shen2025} and $\gamma_{d,\phi}/2\pi = 3~\mathrm{kHz}$~\cite{Ren2022}, and a low temperature $T = 10~\mathrm{mK}$. The results show that the discrepancy between the dynamics predicted by the two Hamiltonians becomes smaller for a larger $\Omega_1$ (thus a larger $|\delta_L| =2\Omega_1$).
	Although a demanding condition $|\delta_L| \gg \left\{\Omega_2,\,g_{\mathrm{q}1}/2,\,g_{\mathrm{q}2}/2,\,\delta_1,\,\delta_2\right \}$ was used for deriving the Hamiltonian~\eqref{Hqm_pp}, numerical results indicate that the condition of  $|\delta_L| \gg \Omega_2$ can be relaxed, since the red curve is a good approximation even for a relatively small ratio $2\Omega_1/\Omega_2  \approx  3.2$. 
	A larger $\Omega_1$ leads to a better approximation of the effective Hamiltonian. However, to avoid using a too strong drive power, the results for $\Omega_1/2\pi>150$ MHz are not shown.  With $\Omega_1/2\pi = 150~\mathrm{MHz}$, we achieve a maximum entanglement $E_N \approx 0.68$ at an optimal time $t \approx 430~\mathrm{ns}$, with only a $4\%$ deviation from that obtained using the effective Hamiltonian, implying that the effective Hamiltonian~\eqref{Heff} is a good approximation under the parameters used in Fig.~\ref{fig2}. 
	{Furthermore, in Appendix~\ref{appendixB} we study the impact of the cavity mode on our two models after eliminating the cavity, i.e., the Hamiltonians~\eqref{Ham_f1} and~\eqref{Heff}.  Numerical results indicate that a too large $\Omega_1$ weakens the validity of the adiabatic elimination and reduces the maximum entanglement that can be achieved, and for a moderate drive strength, e.g., $\Omega_1/2\pi = 120$, the adiabatic elimination of the cavity remains a good approximation under the parameters we used. }
	
\begin{figure}[t]
	\centering
	\includegraphics[width=0.85\linewidth, keepaspectratio]{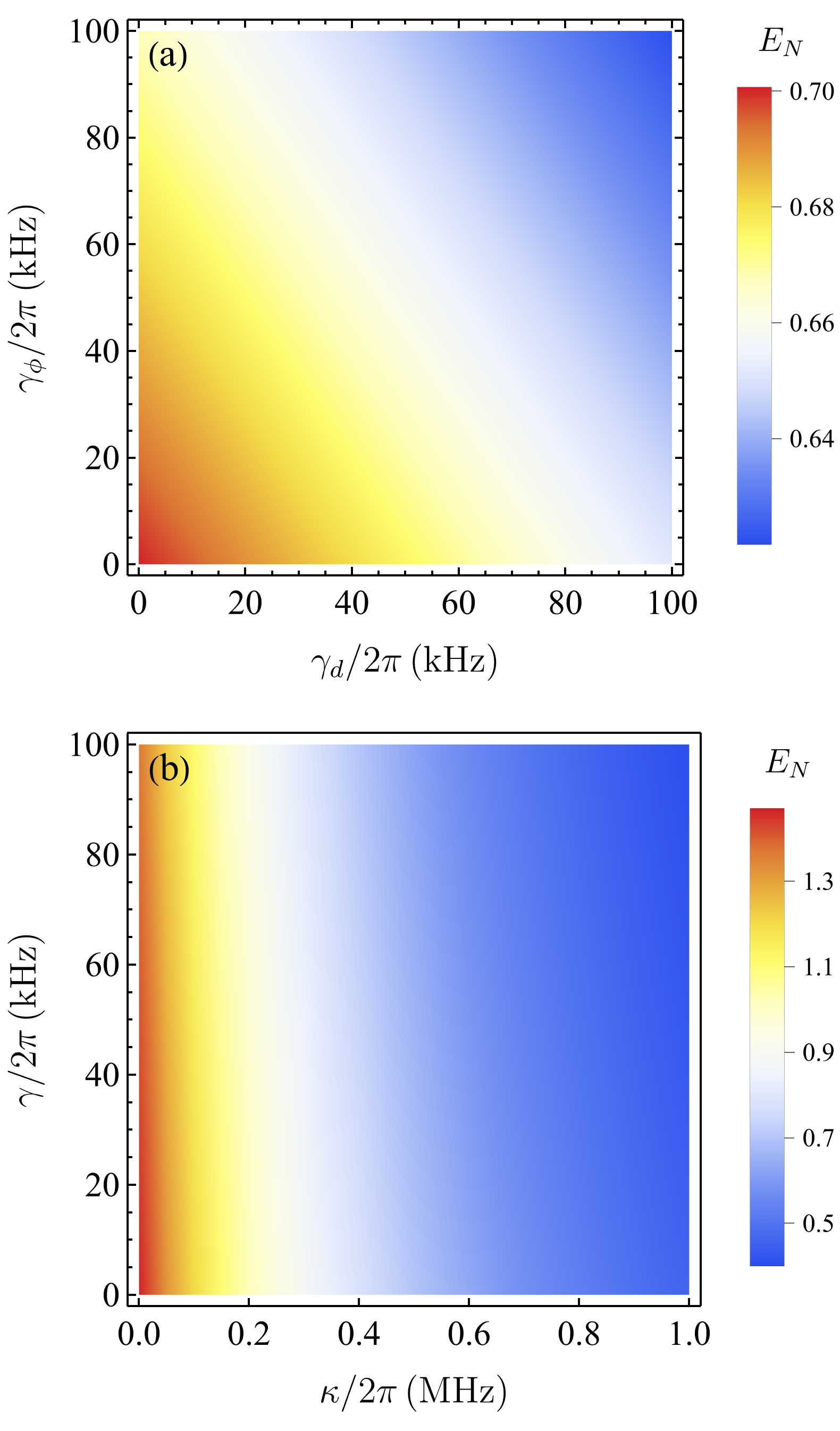}
	\caption{Maximum magnon entanglement as a function of (a) the qubit dissipation rate $\gamma_d$ and dephasing rate $\gamma_\phi$ at a fixed magnon decay rate $\kappa_{1,2}/2\pi = 0.5$ MHz; and (b) the magnon and qubit dissipation rates $\kappa \equiv \kappa_{1,2}$ and $\gamma  \equiv \gamma_{d,\phi}$. The other parameters are those as in Fig.~\ref{fig2}.}
	\label{fig3}
\end{figure}
	
		In Fig.~\ref{fig3}, we plot the maximum entanglement achieved at the optimal time versus the dissipation and dephasing rates of the system. Figure~\ref{fig3}(a) is plotted versus the dissipation and dephasing rates $\gamma_d$ and $\gamma_{\phi}$ of the qubit at a fixed magnon dissipation rate $\kappa_{1,2}/2\pi = 0.5~\text{MHz}$~\cite{Shen2025}. The figure is obtained by numerically solving the master equation in Eq.~\eqref{EME} and the maximum entanglement is extracted from the time-evolved density matrix. It shows that both the qubit dissipation and dephasing rates affect the magnon entanglement almost equally. 
	Figure~\ref{fig3}(b) shows the maximum entanglement versus the magnon and qubit dissipation rates $\kappa_{1} = \kappa_{2} \equiv \kappa$ and $\gamma_d = \gamma_\phi \equiv \gamma$.  {It reveals that the magnon entanglement is mainly limited by the relatively large magnon dissipation rate in the current experiment, which is hardly below 1 MHz.  The dissipation rate can in principle be greatly reduced via, e.g., increasing the purity of the YIG crystal~\cite{Sci.Adv.12.eaee2344}.} 
		
\begin{figure}[t]
		\centering
		\includegraphics[width=0.95\linewidth, keepaspectratio]{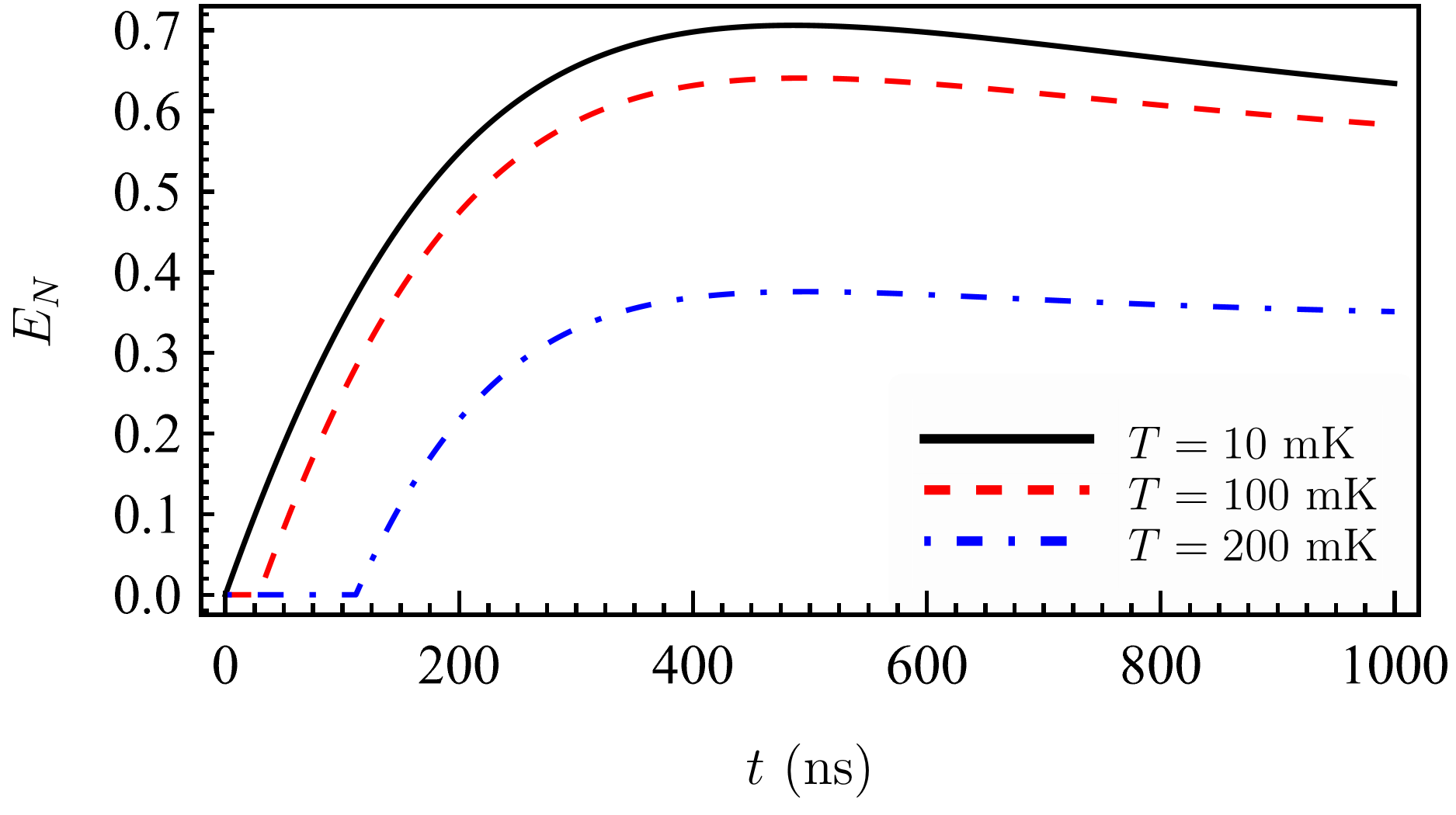}
		\caption{Magnon entanglement as a function of time for different bath temperatures. The parameters are identical to those in Fig.~\ref{fig2}.}
		\label{fig4}
	\end{figure}

	Figure~\ref{fig4} studies the effects of bath temperature on the dynamical entanglement. Note that in this situation the initial state of the system is assumed in $\rho(0) = \ket{g}\bra{g}\otimes \rho_{\text{th}1}\otimes \rho_{\text{th}2}$, where the qubit remains in the ground state while the magnon modes are in the thermal state $\rho_{\text{th}k} = \sum_{n=0}^{\infty}\bar{n}_k^n \ket{n}\bra{n}/(\bar{n}_k+1)^n$ ($k=1,2$). 
	It reveals two main effects as the temperature rises: the maximum entanglement that can be achieved reduces; and it takes longer time for the system to establish the entanglement.  The effect of temperature remains relatively mild for $T < 100$ mK, while temperatures (much) higher than 100 mK significantly reduce the entanglement.

At last, we discuss how to verify the generated entanglement between the two magnon modes. The detection is nontrivial since the system involves two magnon modes and a single qubit, while the magnon state is typically read out via the qubit~\cite{Xuda2023,Xuda2026}. The entanglement can be verified by performing the quantum state tomography of the two magnon modes via measuring their joint Wigner function~\cite{science.aaf2941,Wollack2022}	
	\begin{equation}\label{wigner}
		W(\alpha_1,\alpha_2) = \frac{4}{\pi^2}\mathrm{Tr}\left[D_2(\alpha_2)D_1(\alpha_1)\rho_{12}D_1^\dagger(\alpha_1)D_2^\dagger(\alpha_2)\hat{P} \right],
	\end{equation}
	where $D_k(\alpha_k) = \exp(\alpha_k m_k^\dagger - \alpha_k^* m_k)$ is the displacement operator for the $k$th magnon mode with $\alpha_k \in \mathbb{C}$, and $\hat{P} = e^{i\pi(m_1^\dagger m_1 + m_2^\dagger m_2)}$ is the joint magnon-number parity operator. 
		It indicates that the joint Wigner function can be achieved by measuring the expectation value of the joint parity operator following independent displacements of the two magnon modes. The joint parity operator can be realized via measuring the frequency shift of the qubit induced by the qubit--magnon dispersive interaction~\cite{Lachance2017} and the dispersive shifts caused by the two magnon modes are usually distinguishable~\cite{Wollack2022,science.aaf2941}. To be specific, when the magnon entangled state is prepared, we switch off the two-tone driving field and drive each magnon mode with a resonant microwave field to implement the displacement operation $D_k(\alpha_k)$. We then tune the qubit frequency to a far-detuned point via, e.g., the Autler--Townes effect~\cite{Xuda2023,Xuda2026}, to work in the dispersive regime. For example, for a tuned qubit frequency $\omega_{Q}/2\pi = 5.961~\mathrm{GHz}$, the dispersive shift caused by the first (second) magnon mode is $\chi_{1(2)} \,{\approx}\,\, g_{{q'}_{1(2)}}^2/(\omega_{q} \,{-}\, \omega_{1(2)}) \,{\approx} \, 2\pi \times 0.92$~(0.44)~MHz. A subsequent Ramsey measurement of the qubit yields the joint magnon-number distribution $P_{\alpha_1,\alpha_2}(M_1,M_2) = \langle M_1,M_2 | D_2(\alpha_2)D_1(\alpha_1)\rho_{12} D_1^\dagger(\alpha_1)D_2^\dagger(\alpha_2) | M_1,M_2\rangle$~\cite{Wollack2022}, where $| M_k\rangle$ denotes the $M$-magnon state associated with the $k$th magnon mode. Consequently, the joint Wigner function can be achieved via $W(\alpha_1,\alpha_2) = \frac{4}{\pi^2} P(\alpha_1,\alpha_2)$, with the joint displaced parity $P(\alpha_1,\alpha_2) = \sum_{M_1,M_2}(-1)^{M_1+M_2} P_{\alpha_1,\alpha_2}(M_1,M_2)$.
	
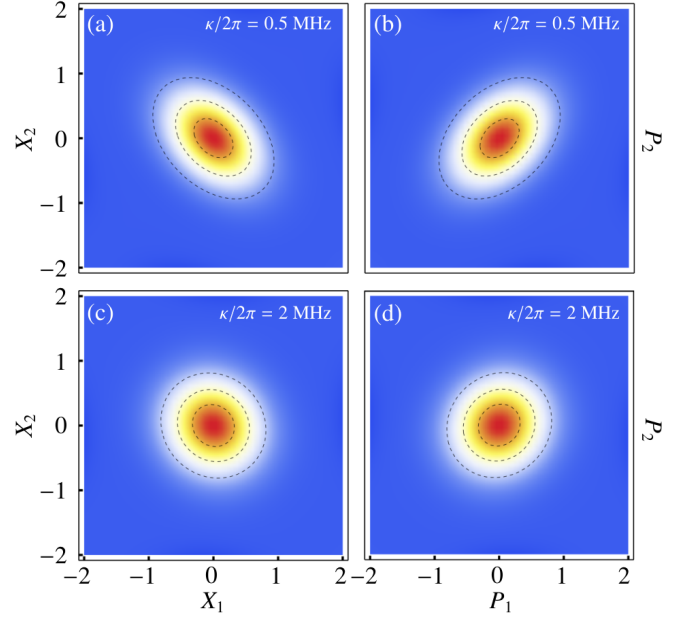
\begin{figure}[tbp]
	\centering
	
	\begin{tikzpicture}
		\node[anchor=south west, inner sep=0] (img)
		at (0,0) {\includegraphics[width=0.88\linewidth, keepaspectratio]{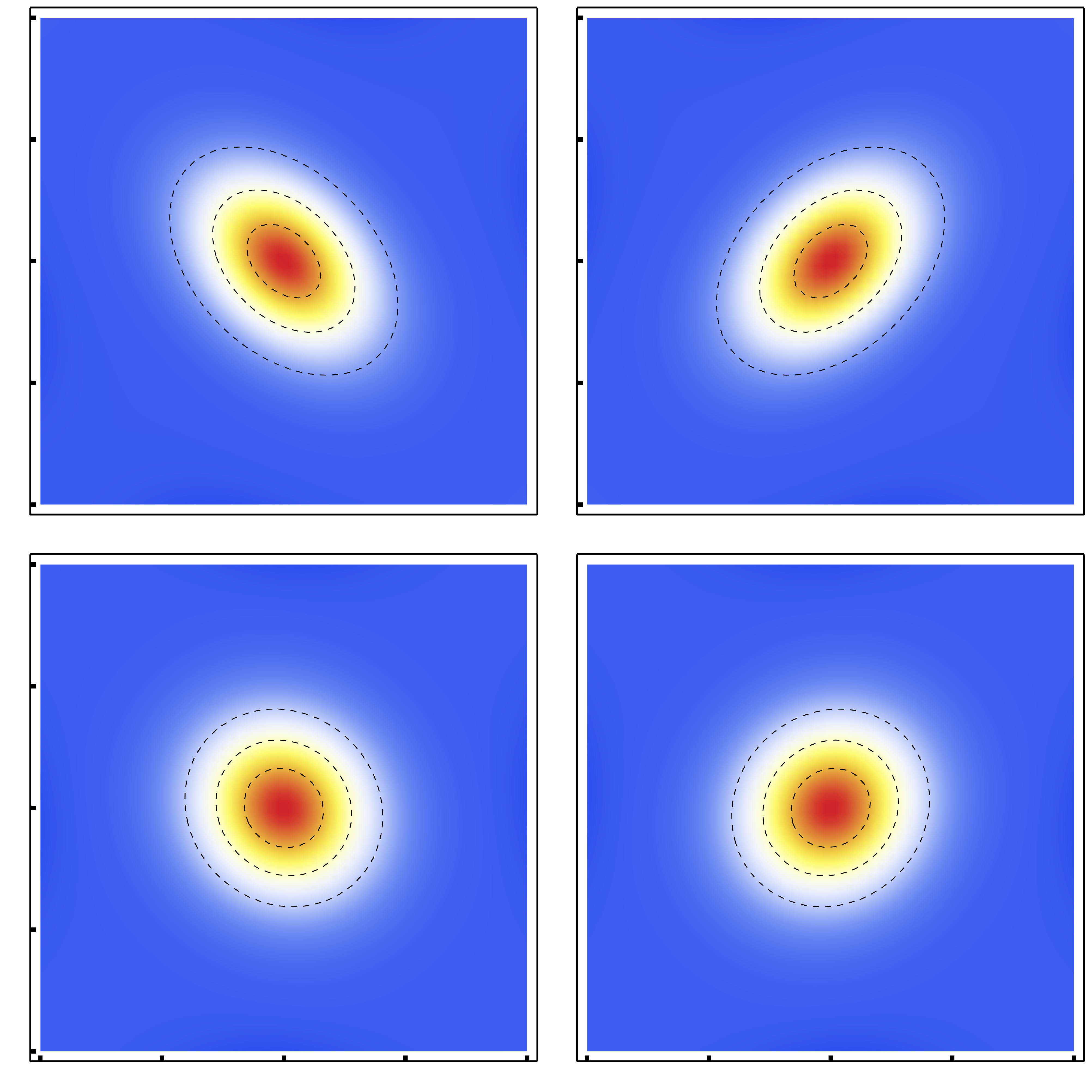}};	
		\begin{scope}[x={(img.south east)}, y={(img.north west)}]
			
			\node[anchor=north east, text=white][font=\scriptsize] at (0.49,0.982) {$\kappa/2\pi = 0.5~\text{MHz}$};
			\node[anchor=north east, text=white][font=\scriptsize] at (0.98,0.982) {$\kappa/2\pi = 0.5~\text{MHz}$};
			\node[anchor=north east, text=white][font=\scriptsize] at (0.49,0.487) {$\kappa/2\pi = 2~\text{MHz}$};
			\node[anchor=north east, text=white][font=\scriptsize] at (0.98,0.487) {$\kappa/2\pi = 2~\text{MHz}$};
			
			\node[anchor=north east, text=white] at (0.115,0.985) {(a)};
			\node[anchor=north east, text=white] at (0.605,0.985) {(b)};
			\node[anchor=north east, text=white] at (0.115,0.487) {(c)};
			\node[anchor=north east, text=white] at (0.605,0.487) {(d)};
			
			\node[anchor=east, rotate=90] at (-0.045,0.796) {$X_2$};			
			\node[anchor=east, rotate=90] at (-0.045,0.31) {$X_2$};
			
			\node[anchor=east] at (0.03,0.979) {$2$};
			\node[anchor=east] at (0.03,0.868) {$1$};
			\node[anchor=east] at (0.03,0.76) {$0$};		
			\node[anchor=east] at (0.03,0.65) {$-1$};
			\node[anchor=east] at (0.03,0.54) {$-2$};
			
			\node[anchor=east] at (0.03,0.488) {$2$};
			\node[anchor=east] at (0.03,0.373) {$1$};
			\node[anchor=east] at (0.03,0.263) {$0$};		
			\node[anchor=east] at (0.03,0.15) {$-1$};
			\node[anchor=east] at (0.03,0.04) {$-2$};
			
			\node[anchor=east] at (0.31,-0.046) {$X_1$};			
			\node[anchor=east] at (0.8,-0.046) {$P_1$};
			
			\node[anchor=east] at (0.06,0.0) {$-2$};
			\node[anchor=east] at (0.19,0.0) {$-1$};
			\node[anchor=east] at (0.288,0.0) {$0$};		
			\node[anchor=east] at (0.397,0.0) {$1$};
			\node[anchor=east] at (0.508,0.0) {$2$};
			
			\node[anchor=east] at (0.568,0.0) {$-2$};
			\node[anchor=east] at (0.682,0.0) {$-1$};
			\node[anchor=east] at (0.788,0.0) {$0$};	
			\node[anchor=east] at (0.897,0.0) {$1$};
			\node[anchor=east] at (1.008,0.0) {$2$};
			
			\node[anchor=east, rotate=-90] at (1.02,0.72) {$P_2$};			
			\node[anchor=east, rotate=-90] at (1.02,0.23) {$P_2$};
			
		\end{scope}
	\end{tikzpicture}
	
	\caption{Joint Wigner function $W(X_1,X_2)$ in (a) and (c) and $W(P_1,P_2)$ in (b) and (d) of two magnon modes at the optimal time giving the maximum entanglement. In (a)-(b), $\kappa/2\pi = 0.5~\mathrm{MHz}$, and (c)-(d) $\kappa/2\pi = 2~\mathrm{MHz}$. The other parameters are identical to those in Fig.~\ref{fig2}.}
	\label{fig5}
	
\end{figure}

	 In Fig.~\ref{fig5}, we plot the joint Wigner functions $W(X_1, X_2)$ and $W(P_1, P_2)$ of the two magnon modes, corresponding to the maximum entanglement achieved at the optimal time, with {$X_k = (m_k+m^\dagger_k)/\sqrt{2}$ and $P_k = (m_k-m^\dagger_k)/\sqrt{2}i$} being the two quadratures of the $k$th magnon mode. 
	 In Fig.~\ref{fig5}(a)-(b), a small magnon decay rate $\kappa/2\pi = 0.5~\mathrm{MHz}$ is used, leading to considerable entanglement with strong two-mode squeezing, while a much larger decay rate $\kappa/2\pi = 2.0~\mathrm{MHz}$ is used in Fig.~\ref{fig5}(c)-(d), resulting in weak entanglement with feeble two-mode squeezing. 

	\section{Conclusions}\label{sec:4}

In conclusion, we have proposed an experimentally feasible scheme for generating macroscopic entanglement between two magnon modes of two YIG spheres in a hybrid cavity--magnon--qubit system. By adiabatically eliminating the cavity mode, properly choosing the magnon and qubit frequencies, and precisely controlling the frequencies and strengths of the two driving fields applied to the qubit, an effective magnonic two-mode squeezing interaction is induced, leading to strong entanglement of two magnon modes under fully feasible parameters.    We further provide a complete detection scheme for verifying the magnon entanglement via measuring the joint Wigner function exploiting the qubit--magnon dispersive coupling. Our results indicate that quantum entanglement between truly macroscopic objects, such as YIG spheres with the diameter ranging from 0.5 mm to 1 mm~\cite{Tabuchi2015,Lachance2017,Lachance2020,Xuda2023,Xuda2026}, is promising to be demonstrated in the near future, and the cavity--magnon--qubit system is an ideal platform for preparing various macroscopic quantum states~\cite{Lachance2017,Lachance2020,Xuda2023,Xuda2026}.

	\begin{acknowledgments}
		 We thank Qi Guo for useful feedback on the manuscript. This work was supported by Zhejiang Provincial Natural Science Foundation of China (LR25A050001), National Natural Science Foundation of China (12474365, 92265202), National Key Research and Development Program of China (2022YFA1405200, 2024YFA1408900), and Fujian Provincial Natural Science Foundation of China (2025J01658).
	\end{acknowledgments}

	\appendix

	\section{Derivation of the effective Hamiltonian~\eqref{Heff0}\label{appendix}}

To derive Eq.~\eqref{Heff0}, we rewrite the interaction Hamiltonian~\eqref{HI}
in the form $H_5 = \sum_{k=1}^{3}[h_k\,e^{-i\tilde\omega_{k}t}
+ h_k^\dagger e^{+i\tilde\omega_{k}t}]$, where
$h_1 = g_{12}m_1m_2^\dagger$,
$h_2 = (g_{\mathrm{q}1}m_1+g_{\mathrm{q}2}m_2^\dagger)\sigma_-/2$,
$h_3 = (g_{\mathrm{q}1}m_1^\dagger+g_{\mathrm{q}2}m_2)\sigma_-/2$,
and the corresponding frequencies are
$\tilde\omega_{1} = 2\delta_m$,
$\tilde\omega_{2} = \Omega_2+\delta_m$,
$\tilde\omega_{3} = \Omega_2-\delta_m$.
Under the condition $\Omega_2>\delta_m$ and
$\{\Omega_2\pm\delta_m,\,2\delta_m\}\gg\{2g_{12},\,g_{\mathrm{q}1},\,g_{\mathrm{q}2}\}/2$,
the effective Hamiltonian is given by the James--Jerke
formula~\cite{James2000,James2007}
\begin{equation}\label{Heff_JJ}
	H_{\mathrm{eff}}
	= \sum_{m,n=1}^{3}
	\frac{1}{\bar{\tilde\omega}_{m,n}}
	\bigl[h_m^\dagger,\,h_n\bigr]\,
	e^{i(\tilde\omega_{m}-\tilde\omega_{n})t},
\end{equation}
where $1/\bar{\tilde\omega}_{m,n} \equiv (1/\tilde\omega_{m}+1/\tilde\omega_{n})/2$.
Expanding Eq.~\eqref{Heff_JJ} over $m,n\in\{1,2,3\}$, we have
\begin{align}
	H_{\mathrm{eff}} = H_{\mathrm{eff,1}} + H_{\mathrm{eff,2}},
\end{align}
where
\begin{align}\label{Heff_expand}
	H_{\mathrm{eff,1}}
	=\;&
	\frac{1}{\tilde\omega_{1}}\bigl[h_1^\dagger,h_1\bigr]
	+\frac{1}{\tilde\omega_{2}}\bigl[h_2^\dagger,h_2\bigr]
	+\frac{1}{\tilde\omega_{3}}\bigl[h_3^\dagger,h_3\bigr],
	\\
	H_{\mathrm{eff,2}}
	=\;&
	\frac{1}{\bar{\tilde\omega}_{12}}\bigl[h_1^\dagger,h_2\bigr]
	e^{i(\tilde\omega_{1}-\tilde\omega_{2})t}
	+\frac{1}{\bar{\tilde\omega}_{13}}\bigl[h_1^\dagger,h_3\bigr]
	e^{i(\tilde\omega_{1}-\tilde\omega_{3})t}
	\nonumber\\
	&+\frac{1}{\bar{\tilde\omega}_{23}}\bigl[h_2^\dagger,h_3\bigr]
	e^{i(\tilde\omega_{2}-\tilde\omega_{3})t}
	+\mathrm{H.c.}
\end{align}
Using $[m_j,m_k^\dagger]=\delta_{jk}$, $n_j\equiv m_j^\dagger m_j$,
and $\sigma_\pm\sigma_\mp=(1\pm\sigma_z)/2$, $H_{\mathrm{eff,1}}$ evaluates to
\begin{align}\label{Heff1}
	H_{\mathrm{eff,1}}
	=&\ \frac{g_{12}^2}{2\delta_m}(n_1-n_2)
	\nonumber\\
	+&\ \frac{1}{4(\Omega_2+\delta_m)}
	\Bigl[\bigl(g_{\mathrm{q}1}^2 n_1+g_{\mathrm{q}2}^2 n_2\bigr)
	+g_{\mathrm{q}1}g_{\mathrm{q}2}
	\bigl(m_1^\dagger m_2^\dagger+m_1 m_2\bigr)\Bigr]\sigma_z
	\nonumber\\
	+&\ \frac{1}{4(\Omega_2-\delta_m)}
	\Bigl[\bigl(g_{\mathrm{q}1}^2 n_1+g_{\mathrm{q}2}^2 n_2\bigr)
	+g_{\mathrm{q}1}g_{\mathrm{q}2}
	\bigl(m_1^\dagger m_2^\dagger+m_1 m_2\bigr)\Bigr]\sigma_z.
\end{align}
Substituting $h_1$, $h_2$, $h_3$ into $H_{\mathrm{eff,2}}$ yields
\begin{align}\label{Heff2}
	H_{\mathrm{eff,2}}
	=&\ \xi_1\bigl[g_{12}m_1^\dagger m_2,\,
	(g_{\mathrm{q}1}m_1+g_{\mathrm{q}2}m_2^\dagger)\sigma_-\bigr]
	e^{-i(\Omega_2-\delta_m)t}
	\nonumber\\
	&+\xi_2\bigl[g_{12}m_1^\dagger m_2,\,
	(g_{\mathrm{q}1}m_1^\dagger+g_{\mathrm{q}2}m_2)\sigma_-\bigr]
	e^{-i(\Omega_2-3\delta_m)t}
	\nonumber\\
	&+\xi_3\bigl[(g_{\mathrm{q}1}m_1+g_{\mathrm{q}2}m_2^\dagger)\sigma^+,\,
	(g_{\mathrm{q}1}m_1+g_{\mathrm{q}2}m_2^\dagger)\sigma_-\bigr]
	e^{-i2\delta_m t}
	\nonumber\\
	&+\mathrm{H.c.},
\end{align}
where $\xi_1=(\Omega_2+3\delta_m)/[8\delta_m(\Omega_2+\delta_m)]$,
$\xi_2=(\Omega_2+\delta_m)/[8\delta_m(\Omega_2-\delta_m)]$,
and $\xi_3=\Omega_2/[4(\Omega_2^2-\delta_m^2)]$.
Collecting Eqs.~\eqref{Heff1} and~\eqref{Heff2} reproduces Eq.~\eqref{Heff0}.
When the condition $\{\Omega_2-\delta_m,\,|\Omega_2-3\delta_m|,\,2\delta_m\}	\gg\{\xi_1 g_{12}g_{\mathrm{q}k},\,\xi_2 g_{12}g_{\mathrm{q}k},\,	\xi_3 g_{\mathrm{q}k}g_{\mathrm{q}l}\}$ ($k,l\in\{1,2\}$) is satisfied,	all three oscillating terms in $H_{\mathrm{eff,2}}$ are rapidly varying	on the timescale of the slow effective dynamics and can be neglected	under a RWA, leaving $H_{\mathrm{eff}}\approx H_{\mathrm{eff,1}}$.	Combining the second and third lines of Eq.~\eqref{Heff1} and introducing	$r_g=\bigl[(\Omega_2-\delta_m)^{-1}+(\Omega_2+\delta_m)^{-1}\bigr]/4$	and $g_{\mathrm{eff}}=r_g g_{\mathrm{q}1}g_{\mathrm{q}2}$	then yields Eq.~\eqref{Heff}.

	\begin{figure}[t]
			\centering
			\includegraphics[width=0.95\linewidth,keepaspectratio]{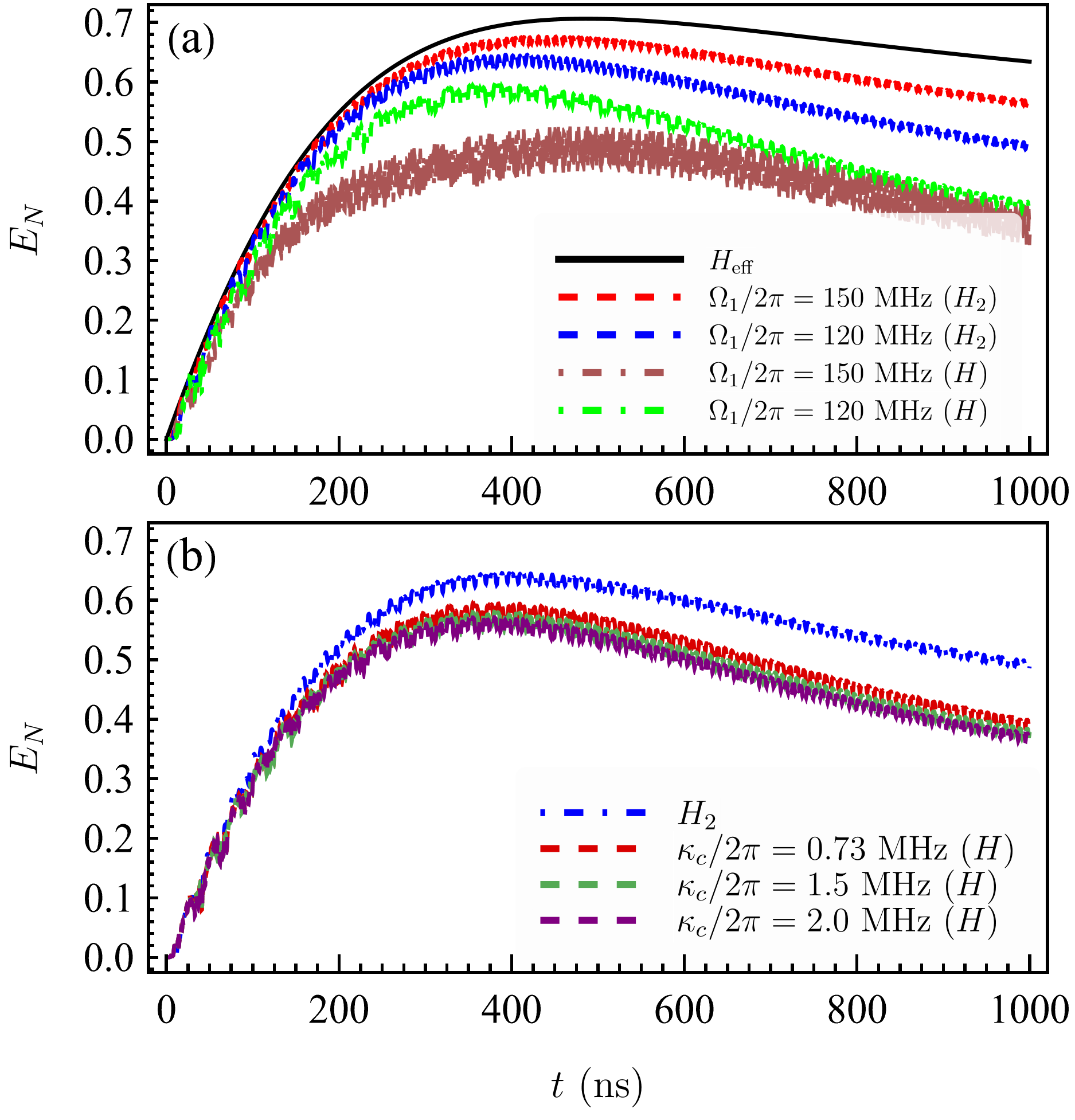}
			\caption{Entanglement between two magnon modes as a function of time. (a) Comparison among different models: the black solid curve represents the dynamics governed by the effective Hamiltonian~\eqref{Heff}; {the red and blue dashed curves} correspond to the magnon--qubit Hamiltonian~\eqref{Ham_f1} for $\Omega_1/2\pi = 120\text{ MHz}$ and $150\text{ MHz}$, respectively; {the green and brown dot-dashed curves} are obtained from the full Hamiltonian~\eqref{Ham_ff} incorporating the cavity mode, with a cavity decay rate of $\kappa_c/2\pi = 0.73\text{ MHz}$. (b) Influence of the cavity decay rate $\kappa_c$ on the entanglement dynamics for $\Omega_1/2\pi = 120\text{ MHz}$. Other system parameters are identical to those used in Fig.~\ref{fig2}.}
			\label{figA12}
		\end{figure}

	\section{Effect of the cavity mode on magnon entanglement\label{appendixB}}

		Our derivation starts from the effective Hamiltonian~\eqref{Ham_full} of the magnon--qubit system after adiabatically eliminating the cavity mode. To perform a benchmark comparison, here we perform a numerical simulation based on the master equation of the original full Hamiltonian~\eqref{Ham_ff} including the cavity, which is
		\begin{align}\label{ME1}  
			\frac{d\rho}{dt} =& - i [H, \rho] + \sum_{k=1}^{2} \left( \frac{\kappa_k(\bar{n}_{k} + 1)}{2} \mathcal{L} [m_k]\rho + \frac{\kappa_k \bar{n}_{k}}{2} \mathcal{L} [m_k^\dag]\rho \right) \nonumber\\
			&+ \frac{\kappa_c(\bar{n}_{c} + 1)}{2} \mathcal{L} [c]\rho + \frac{\kappa_c \bar{n}_{c}}{2} \mathcal{L} [c^\dag]\rho \nonumber\\
			&+ \frac{\gamma_d (\bar{n}_{q} + 1)}{2} \mathcal{L} [\sigma_-]\rho + \frac{\gamma_d \bar{n}_{q}}{2} \mathcal{L} [\sigma_+]\rho + \frac{\gamma_{\phi}}{4} \mathcal{L} [\sigma_z]\rho.
		\end{align}	
		Here, the Hamiltonian $H$ is given in Eq.~\eqref{Ham_ff}, $\kappa_c$ is the cavity decay rate, and $\bar{n}_c$ is the mean thermal photon number. 	
		
	In Fig.~\ref{figA12}(a), we compare the numerical results obtained by solving the master equations in Eqs.~\eqref{EME},~\eqref{ME}, and~\eqref{ME1}, which correspond to the Hamiltonians~\eqref{Heff}, \eqref{Ham_f1}, and~\eqref{Ham_ff}, respectively.  The results confirm that the effective Hamiltonian $H_{\text{eff}}$ in Eq.~\eqref{Heff} predominantly captures the system dynamics. Figure~\ref{fig2} indicates that increasing the Rabi frequency $\Omega_1$ enhances the entanglement. {However, when the cavity is taken into account, there is a trade-off effect regarding $\Omega_1$: if $\Omega_1$ is too small, the large-detuning condition $\delta_L = -2\Omega_1$ for the RWA is no longer well satisfied; On the contrary, a larger $\Omega_1$ may lead the transition between the dressed states $\ket{-}\ket{n}$ and $\ket{+}\ket{n+1}$ closer to the cavity frequency,} which weakens the validity of the adiabatic elimination and may reduce the maximum entanglement that can be achieved.
	For instance, the full master equation Eq.~\eqref{ME1} yields the maximum entanglement of {$E_N = 0.59$ ($0.51$) for $\Omega_1/2\pi = 120$ MHz (150 MHz).}
	For $\Omega_1/2\pi = 120$ MHz, the effective Hamiltonian \eqref{Heff} remains a fair approximation with a deviation of {$\sim 8.6\%$} from the maximum entanglement achieved using the reduced magnon--qubit Hamiltonian \eqref{Ham_f1} and {$\sim 18.7\%$} from that using the full Hamiltonian~\eqref{Ham_ff}. Figure 6(b) reveals a {slight}  impact of the cavity decay rate on the deviation between the two models ($H$ and $H_2$) before and after adiabatically eliminating the cavity. For $\kappa_c/2\pi = 0.73$ MHz reported in the experiment~\cite{Xuda2026}, the full Hamiltonian~\eqref{Ham_ff} yields a maximum entanglement {$E_N=0.59$, which is slightly smaller than the maximum value $E_N=0.64$ achieved using the Hamiltonian \eqref{Ham_f1}.}

\bibliography{references}
	
\end{document}